%% file: 0main.tex
\documentclass[conference,a4paper]{IEEEtran}
\IEEEoverridecommandlockouts
\usepackage{cite}
\usepackage[
top    = 1.84cm,
bottom = 4.05cm,
left   = 1.43cm,
right  = 1.43cm]{geometry}
\usepackage{amsmath,amssymb,amsfonts}
\usepackage{algorithmic}
\usepackage{lipsum,adjustbox}
\usepackage{verbatim}
\usepackage{graphics}
\usepackage{stackengine}
\usepackage{pgfplots}
\pgfplotsset{width=7cm,compat=1.8}
\usepackage{multirow}
\usepackage{soul}
\usepackage{amssymb}
\usepackage{amsmath}
\usepackage{algorithm}
\usepackage{algorithmic}
\usepackage{graphicx}
\usepackage{subcaption}
\usepackage{textcomp}
\usepackage{booktabs}
\usepackage{stfloats}
\usepackage{scalefnt}
\usepackage{mathrsfs}       
\usepackage[nolist]{acronym}
\input{commands}
\input{acronyms.tex}
\usepackage[utf8]{inputenc}
\usepackage[english]{babel}
\usepackage{amsthm}
\usepackage{commath}

\begin{document}

\title{Deep Learning Based Channel Estimation in High Mobility Communications Using Bi-RNN Networks
}

\author{\IEEEauthorblockA{Abdul Karim Gizzini\IEEEauthorrefmark{1},
Marwa Chafii\IEEEauthorrefmark{2}\IEEEauthorrefmark{3},
}

\IEEEauthorblockA{\IEEEauthorrefmark{1}ETIS, UMR8051,
CY Cergy Paris Université, ENSEA, CNRS, France\\
\IEEEauthorrefmark{2} Engineering Division, New York University (NYU), Abu Dhabi 129188, UAE \\
\IEEEauthorrefmark{3} NYU WIRELESS, NYU Tandon School of Engineering, Brooklyn, 11201, NY\\
abdulkarim.gizzini@ensea.fr, marwa.chafii@nyu.edu }}

\maketitle

\input{1abstract}
\input{2introduction}
\input{3system_model}

\input{5proposed_estimator}

\input{6simulation_results}
\input{7conclusions}
\bibliographystyle{IEEEtran}
\bibliography{ref}
\end{document}

%% file: commands.tex
 \usepackage{bm}

\newcommand{\figref}[1]{Figure~\ref{#1}}

\newcommand{\compl}{\mathbb{C}}         

\newcommand{\ma}  [1]{ \bm{#1} } 

\newcommand{\set} [1]{{\mathcal {#1}}} 
\newcommand{\Kon} {\set{K}_{\text{on}}} 
\newcommand{\Ip} {\set{I}_{\text{p}}} 
\newcommand{\Id} {\set{I}_{\text{d}}} 



%% file: acronyms.tex
\begin{acronym}
    \acro{C-ITS}{cooperative intelligent transportation system}
    \acro{RBF}{radial basis function}
    \acro{ANN}{artificial neural network}
\acro{GRU}{gated recurrent unit}
\acro{RNN}{Recurrent neural network}
\acro{Bi-RNN}{bi-directional recurrent neural network}
\acro{Bi}{bi-directional}
\acro{FNN}{feed-forward neural network}
    \acro{CNN}{convolutional neural network}
    \acro{TS-ChannelNet}{Temporal spectral ChannelNet}
    \acro{TDR}{transmission data rate}
    \acro{LSTM}{long short-term memory}
    \acro{ALS}{Accurate LS}
    \acro{SLS}{simple LS}
    \acro{ChannelNet}{channel network}
    \acro{ADD-TT}{average decision-directed with time truncation}
    \acro{WI}{weighted interpolation}
    \acro{DD}{decision-directed}
    \acro{SR-ConvLSTM}{super-resolution convolutional long short-term memory}
    \acro{SR-CNN}{super resolution CNN}
    \acro{DN-CNN}{denoising CNN}
    \acro{V2V}{vehicle-to-vehicle}
    \acro{V2I}{vehicle-to-infrastructure}
    \acro{VTV}{vehicle-to-vehicle}
    \acro{RTV}{roadside-to-vehicle}
    \acro{DPA}{data-pilot aided}
    \acro{STA}{spectral temporal averaging}
    \acro{M-STA}{modified STA}
    \acro{LMMSE}{linear minimum mean-squared error}
    \acro{M-CDP}{modified CDP}
    \acro{M-TRFI}{modified TRFI}
    \acro{SNR}{signal-to-noise ratio}
    \acro{CDP}{Constructed data pilots}
    \acro{DFT}{discrete Fourier transform}
    \acro{T-DFT}{truncated discrete Fourier transform}
    \acro{TA-MDFT}{temporal averaging M-DFT}
    \acro{TA-DFT}{TA-DFT}
    \acro{STS}{short training symbols}
    \acro{LTS}{long training symbols}
    \acro{SS}{signal symbol}
    \acro{TDL}{tapped delay line}
    \acro{TRFI}{time domain reliable test frequency domain interpolation}
    \acro{SBS}{symbol-by-symbol}
    \acro{FBF}{frame-by-frame}
    \acro{MMSE-VP}{minimum mean square error using virtual pilots}
    \acro{DL}{deep learning}
    \acro{AE-DNN}{auto-encoder deep neural network}
    \acro{DNN}{deep neural networks} 
    \acro{ISI}{inter-symbol-interference}
    \acro{OFDM}{orthogonal frequency-division multiplexing}
    \acro{LS}{least square}
    \acro{RS}{reliable subcarriers}
    \acro{URS}{unreliable subcarriers}
    \acro{SRNN}{Simple RNN}
    \acro{MMSE}{minimum mean squared error}
    \acro{SoA}{state of the art}
    \acro{NMSE}{normalized mean-squared error}
    \acro{AWGN}{additive white Gaussian noise}
	\acro{BER}{bit error rate}
	\acro{RSU}{road side unit}
	\acro{AE}{auto-encoder}
    \acro{CP}{cyclic prefix}
\acro{PLCP}{physical layer convergence protocol}
\acro{MSE}{mean squared error}
\end{acronym}

%% file: 1abstract.tex
\begin{abstract}

Doubly-selective channel estimation represents a key element in ensuring communication reliability in wireless systems. Due to the impact of multi-path propagation and Doppler interference in dynamic environments, doubly-selective channel estimation becomes challenging. Conventional channel estimation schemes encounter performance degradation in high mobility scenarios due to the usage of limited training pilots. Recently, deep learning (DL) has been utilized for doubly-selective channel estimation, where {\ac{CNN}} networks are employed in the {\ac{FBF}} channel estimation. However, CNN-based estimators require high complexity, making them impractical in real-case scenarios. For this purpose, we overcome this issue by proposing an optimized and robust {\ac{Bi-RNN}} based channel estimator to accurately estimate the doubly-selective channel, especially in high mobility scenarios. The proposed estimator is based on performing end-to-end interpolation using {\ac{GRU}} unit. Extensive numerical experiments demonstrate that the developed Bi-GRU estimator significantly outperforms the recently proposed CNN-based estimators in different mobility scenarios, while substantially reducing the overall computational complexity.
\end{abstract}


%% file: 2introduction.tex
\section{Introduction} \label{introduction}
\IEEEPARstart{T}he recent advances in wireless communications enable high data rates and low latency mobile wireless applications~\cite{ref_marwa6G}. Wireless communications offer mobility to different nodes within the network, however, the mobility feature has a severe negative impact on the communication reliability~\cite{bomfin2021robust}.
In such environment, the wireless channel is said to be doubly-selective, i.e. varies in both time and frequency. This is due to the propagation medium, where the transmitted signals propagate through multiple paths, each having a different power, delay, and Doppler shift effect resulting from the motion of network nodes. Knowing that the accuracy of the estimated channel influences the system performance, since it affects  different operations at the receiver like equalization, demodulation, and decoding. Therefore, ensuring communication reliability using accurate channel estimation is crucial, especially in high mobility scenarios.

The {\ac{SoA}} channel estimation schemes can be categorized into \ac{SBS} estimators where the channel is estimated for each received symbol separately using only the previous and current pilots, and \ac{FBF} estimators where the previous, current and future pilots are employed in the channel estimation for each received symbol~\cite{9813719}. The higher channel estimation accuracy can be achieved by using {\ac{FBF}} estimators, since the channel estimation of each symbol takes advantage from the knowledge of previous, current, and future allocated pilots within the frame. The well-known {\ac{FBF}} estimator is the conventional 2D \ac{LMMSE}~\cite{ref_LMMSE_Computational} where the channel and noise statistics are utilized in the estimation, thus, leading to comparable performance to the ideal case. However, the 2D {\ac{LMMSE}} suffers from high computational complexity making it impractical in real case scenarios. Therefore, there is a  need for robust and low-complexity {\ac{FBF}} estimators.

Recently, {\ac{DL}} algorithms have been integrated into wireless communications physical layer applications~\cite{ref_DL_PHY2} including channel estimation~\cite{ref_DL_Chest1, ref111,ref_STA_DNN,ref_ELS,ref_trfifnn,ref_adaptiveELS}, due to its great success in improving the overall system performance, especially when used on top of low-complexity conventional estimators. {\ac{DL}} algorithms are characterized by robustness, low-complexity, and good generalization ability making the integration of {\ac{DL}} into communication systems beneficial. Motivated by these advantages, {\ac{DL}} algorithms have been integrated into doubly-selective {\ac{FBF}} channel estimation where the initial estimated channel for the whole frame is considered as a 2D low-resolution noisy image and \ac{CNN}-based processing is applied as super-resolution and denoising techniques. 

In~\cite{ref_ChannelNet}, the authors propose a {\ac{CNN}}-based channel estimator denoted as {\ac{ChannelNet}}, that applies {\ac{RBF}} interpolation as initial channel estimation, after that, the {\ac{RBF}} estimated channel is considered as a low resolution image, where {\ac{SR-CNN}} followed by {\ac{DN-CNN}} are integrated on top of the {\ac{RBF}} estimated channel. {\ac{TS-ChannelNet}} has been proposed in~\cite{ref_TS_ChannelNet}, where {\ac{ADD-TT}} interpolation that is based on the demodulation and averaging of each received symbol. After that, {\ac{SR-ConvLSTM}} is used to improve {\ac{ADD-TT}} interpolation accuracy. {\ac{TS-ChannelNet}} estimator outperforms the {\ac{ChannelNet}} estimator especially in low {\ac{SNR}} regions. Moreover, the {\ac{TS-ChannelNet}} estimator has lower computational complexity than the {\ac{ChannelNet}} estimator since only one {\ac{CNN}} network is considered instead of two {\acp{CNN}} as proposed in the {\ac{ChannelNet}} estimator. Nevertheless, {\ac{ChannelNet}} and {\ac{TS-ChannelNet}} suffer from a considerable performance degradation in high mobility scenarios due to the high estimation error of the employed initial estimation. The {\ac{WI}}-{\ac{CNN}} estimators have been proposed in~\cite{WI-CNN} to further improve the performance in high mobility scenarios, where new frame design has been utilized. Moreover, the comb pilots allocation has been replaced by periodically inserting pilot symbols within the transmitted frame according to the mobility scenario. After that, the channel estimation is performed according to three main steps: (\textit{i}) Channel estimation at the inserted pilot symbols employing several {\ac{LS}} estimation schemes. (\textit{ii}) Channel estimation at the remaining data symbols, where {\ac{WI}}  of the estimated channels at the inserted pilot symbols are applied. (\textit{iii}) Integrating optimized {\ac{SR-CNN}} or {\ac{DN-CNN}} as post-processing modules after the {\ac{WI}} estimators. Even though, the WI-CNN estimator improves the overall performance in comparison to the  {\ac{ChannelNet}}, {\ac{TS-ChannelNet}} estimators, but it employs two {\ac{CNN}} networks that are selected based on the mobility scenario. Moreover, {\ac{ChannelNet}}, {\ac{TS-ChannelNet}}, and {\ac{WI}}-{\ac{CNN}} estimators suffer from high computational complexity, due to the huge operations applied by the employed {\ac{CNN}} architectures.

It is worth mentioning that {\ac{CNN}} networks are used basically to alleviate the impact of noise within the input frame, where they improve the resolution as the case of {\ac{SR-CNN}}~\cite{ref_SRCNN}, whereas, {\ac{DN-CNN}}~\cite{ref_DNCNN} works on noise extraction using residual learning~\cite{ref_residual_learning}, then the input frame is subtracted from the extracted noise and the denoised output is obtained.

Motivated by the fact that {\acp{Bi-RNN}} are designed to perform 2D interpolation of unknown data bounded between known data, especially, when working with correlated data~\cite{bi-RNN}. This paper focuses on overcoming the limitations of the {\ac{SoA}} {\ac{CNN}}-based channel estimation schemes, where a {\ac{Bi-RNN}} based channel estimation scheme is proposed. The proposed scheme inherits the adaptive frame design from the {\ac{WI}}-{\ac{CNN}} estimators. It performs the channel estimation at the inserted pilot symbols, after that, it employs a Bi-{\ac{GRU}} as an end-to-end 2D interpolation unit to estimate the channel at the data symbols. Unlike the {\ac{WI}}-\ac{CNN} estimators, where {\ac{WI}} followed by {\ac{SR-CNN}} and {\ac{DN-CNN}} processing in low and high mobility scenarios, respectively, the proposed channel estimator uses the same Bi-GRU unit for all mobility scenarios. Moreover, the 2D interpolation is performed completely by this Bi-GRU unit without the need to any initial estimation. Simulation results show the performance superiority of the proposed Bi-RNN based channel estimation scheme against the {\ac{SoA}} {\ac{CNN}}-based channel estimators while recording an outstanding computational complexity reduction. 



    
    

The remainder of this paper is organized as follows: in Section~\ref{system_model}, the system model is described. The proposed {\ac{Bi-RNN}} based channel estimation scheme is presented in Section~\ref{Proposed_scheme}. In Section~\ref{simulation_results}, the performance evaluation in terms of \ac{BER} as well as the computational complexity analysis for the studied channel estimators are presented and discussed. Finally, the paper is concluded in Section~\ref{conclusions}. 

%% file: 3system_model.tex
\section{System Model} \label{system_model}

Consider a frame consisting of $I = I_{d} + P$ {\ac{OFDM}} symbols, where $P$ denotes the number of inserted pilot symbols and $I_{d}$ refers to the remaining {\ac{OFDM}} data symbols that are preserved for actual data transmission. The $i$-th transmitted frequency-domain {\ac{OFDM}} symbol $\tilde{\ma{x}}_i[k] \in \compl^{K_{\text{on}} \times 1}$ can be expressed as follows
\begin{equation}
   \tilde{\ma{x}}_i[k] = \left\{
            \begin{array}{ll}        
                \tilde{\ma{x}}_{{d}}[k],&\quad i \in \Id. \\
                \tilde{\ma{x}}_{{p}}[k],&\quad i \in \Ip. \\
            \end{array}\right.
\label{eq: xK}
\end{equation}

where $0 \leq k \leq K_{\text{on}} - 1$ and $K_{\text{on}}$ refers to the allocated active subcarriers. $\tilde{\ma{x}}_{{d}}[k]$ and $ \tilde{\ma{x}}_{{p}}[k]$ represent the modulated data symbols and the predefined pilot symbols allocated at $\Id$ and $\Ip$ sets, respectively.
The received frequency-domain {\ac{OFDM}} symbol denoted as $\tilde{\ma{y}}_{{i}}[k] \in \compl^{K_{\text{on}} \times 1}$ is expressed as follows

\begin{equation}
	\begin{split}
		\tilde{\ma{y}}_{{i}}[k] 
		&= \tilde{\ma{h}}_i[k] \tilde{\ma{x}}_i[k] + \tilde{\ma{v}}_i[k],~ k \in \Kon.
	\end{split}            
	\label{eq: system_model}
\end{equation}

$\tilde{\ma{h}}_i[k] \in \compl^{K_{\text{on}} \times 1}$ refers to the frequency response of the doubly-selective channel at the $i$-th {\ac{OFDM}} symbol and $k$-th subcarrier. ${\tilde{\ma{v}}}_i[k]$ signifies the \ac{AWGN} of variance $\sigma^2$. As a matrix form, \eqref{eq: system_model} can be expressed as follows
 
\begin{equation}
\tilde{\ma{Y}}[k,i] = \tilde{\ma{H}}[k,i]  \tilde{\ma{X}}[k,i] + \tilde{\ma{V}}[k,i],~ k \in \Kon,
\label{eq:preamble_freq}
\end{equation}
where $\tilde{\ma{V}}[k,i] \in \compl ^{K_{\text{on}}\times I}$ and $\tilde{\ma{H}} \in \compl ^{K_{\text{on}}\times I}$ denote the {\ac{AWGN}} noise and the doubly-selective frequency response of the channel for all symbols within the transmitted \ac{OFDM} frame, respectively. The received pilot symbols are denoted by $\tilde{\ma{Y}}_{P} = [\tilde{\ma{y}}^{(p)}_{1}, \dots,  \tilde{\ma{y}}^{(p)}_{q}, \dots, \tilde{\ma{y}}^{(p)}_{P}] \in \compl ^{K_{\text{on}}\times P}$. $q$ denotes the inserted pilot symbol index, where $1 \leq q \leq P$.

%% file: 5proposed_estimator.tex
\section{Proposed Bi-RNN based channel estimation schemes} \label{Proposed_scheme}

RNN is a type of \ac{DL} designed to work with sequential data. This sequential data can be in form of time series, text, audio, video etc. Regular RNN uses the previous and the current information in the sequence to produce the current output, whereas, Bi-RNN networks are designed to predict unknown data that are bounded within known data~\cite{bi-RNN}. They are based on making 
the data flows through any RNN unit in both directions forward (past to future), and  backward (future to past).  In regular RNN, the input flows in one direction, whereas, in {\ac{Bi-RNN}} the input flows in both directions to get the advantage of both past and future information. By doing so, the {\ac{Bi-RNN}} network will be able to predict the unknown information in the middle based on their correlation with the known past and future information.
\begin{figure*}[t]
	\centering
	\includegraphics[width=2\columnwidth]{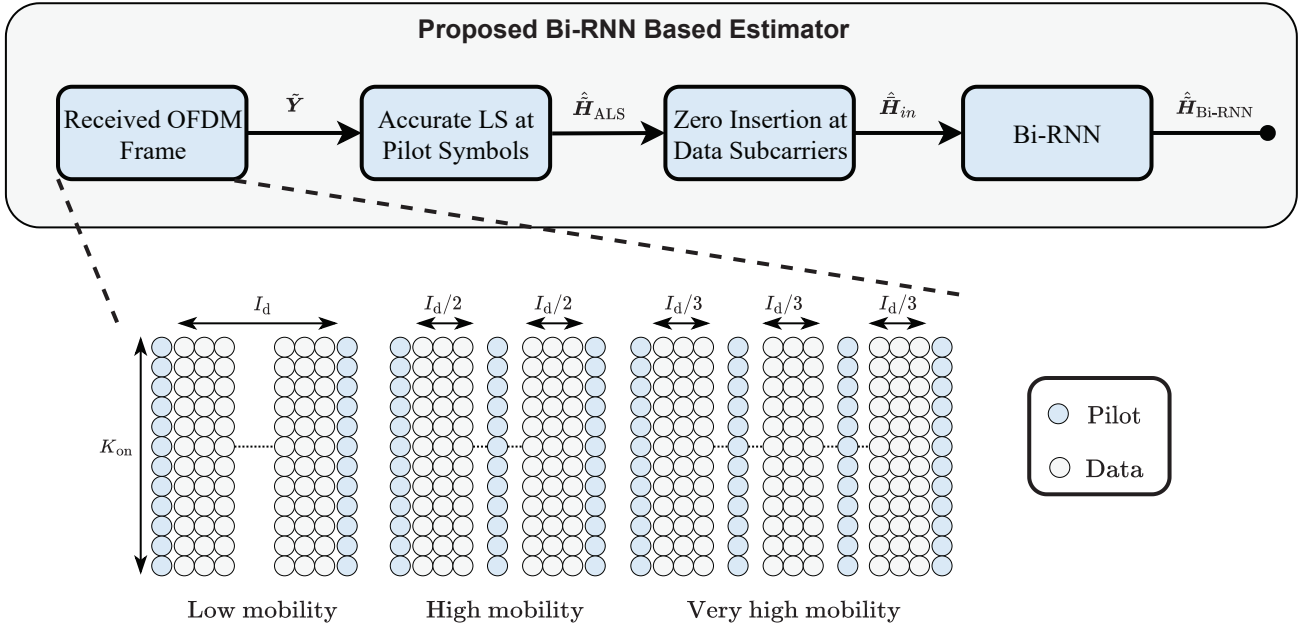}
	\caption{Proposed Bi-RNN based channel estimator block diagram.}
	\label{fig:proposed-bi-rnn}
\end{figure*}
In this context, the proposed Bi-RNN channel estimator aims to utilize the interpolation ability of Bi-RNN networks in the {\ac{FBF}} channel estimation instead of employing high-complexity CNN networks as it is the case in the {\ac{SoA}} {\ac{CNN}}-based channel estimation schemes.


\begin{table}
	\renewcommand{\arraystretch}{1.4}
	\centering
	\caption{Parameters of the proposed Bi-RNN-based channel estimation scheme.}
	\label{tb:LSTM_params}
	\begin{tabular}{l|l}
	\hline
		(Bi-GRU; Hidden size ($Q$)) & (1;32)  \\ \hline
		Activation function              & ReLU $(y= \max(0,x))$                     \\ \hline
		Number of epochs        & 500                                \\ \hline
		Training samples        & 16000                             \\ \hline
		Testing samples        & 2000                             \\ \hline
		Batch size          & 128                                    \\ \hline
		Optimizer       & ADAM                                       \\ \hline
		Loss function      & MSE                                     \\ \hline
	    Training SNR        & 40 dB                                 \\ \hline
	\end{tabular}
\end{table}

The proposed Bi-RNN channel estimation scheme uses Bi-GRU unit and it inherits the adaptive frame design from the WI-CNN estimators as shown in Figure~\ref{fig:proposed-bi-rnn}. We note that the frame structure can be adapted according to the considered scenario. For example, in vehicular communications, the low mobility frame structure is used in urban environments since the car velocity must not exceed 40 Kmphr. Similarly, for the highways environment. Recall that WI-{\ac{CNN}} channel estimation performs WI interpolation at the data symbols, where the initial estimated channels are modeled as a 2D noisy image and {\ac{CNN}} processing is applied to alleviate the impact of noise. However, Bi-RNN performs 2D interpolation at the data symbols using the estimated channel at the pilot symbols without the need for any initial channel estimation at the data symbols. The proposed Bi-RNN channel estimator proceeds as follows

\begin{itemize}
    \item {\ac{ALS}} estimation at the inserted pilot symbols. The {\ac{ALS}} relies on estimating the channel impulse response ${\ma{h}}_{q,L} \in \compl ^{L\times 1}$ at the $q$-th received pilot symbol and then applying {\ac{DFT}} interpolation as follows
    
        \begin{equation}
        \hat{\ma{h}}_{q,L} = \ma{F}_{\text{on}}^{\dagger} \hat{\tilde{\ma{h}}}_{{\text{LS}}_{q}},~  k \in \Kon,
        \label{eq:ALS1}
        \end{equation}
        
        \begin{equation}
        \hat{\tilde{\ma{h}}}_{{\text{ALS}}_{q}} = \ma{F}_{\text{on}} \hat{\ma{h}}_{q,L},  ~ k \in \Kon.
        \label{eq:ALSP}
        \end{equation}
        where $\ma{F}_{\text{on}}^{\dagger} = [(\ma{F}_{\text{on}}^{\text{H}} \ma{F}_{\text{on}})^{-1} \ma{F}_{\text{on}}^{\text{H}}]$ is the pseudo inverse of the scaled DFT matrix $\ma{F}_{\text{on}} \in \compl^{K_{\text{on}} \times L}$. Moreover, $\hat{\tilde{\ma{h}}}_{{\text{LS}}_{q}}[k]$ denotes the {\ac{LS}} channel estimation at the $q$-th pilot symbol that can be calculated in terms of the pre-defined pilot symbol $\tilde{\ma{p}}[k]$ such that  
        
        \begin{equation}
        \hat{\tilde{\ma{h}}}_{{\text{LS}}_{q}}[k] = \frac{\tilde{\ma{y}}^{(p)}_{q}[k]}{\tilde{\ma{p}}[k]}, ~ k \in \Kon.
        \label{eq: SLSP}
        \end{equation}
       
    \item  After that, zero insertion at all the data symbols is applied. The obtained frame $\hat{\tilde{\ma{H}}}_{{\rho}} \in \compl^{K_{\text{on}} \times I}$ is converted to the real-valued domain by vertically stacking the real and imaginary values to get $\hat{\Bar{\ma{H}}}_{{in}} \in \mathbb{R}^{2K_{\text{on}} \times I}$.
    
    \item Bi-RNN end-to-end interpolation, where $\hat{\Bar{\ma{H}}}_{{in}}$ is fed as an input to the optimized Bi-GRU unit. By doing so, the Bi-GRU unit learns the weights of the estimated channels at the OFDM data symbols taking into consideration the influence of the estimated channel at the pilot symbols within the frame. Employing the 2D interpolation using the proposed Bi-GRU unit leads to a considerable performance superiority in comparison with the WI-CNN estimators while recording a significant decrease in the required computational complexity, as shown in Section~\ref{simulation_results}. We note that the proposed Bi-GRU architecture is optimized using the grid search algorithm~{\cite{ref_grid_search}} and trained using the parameters listed in~Table~{\ref{tb:LSTM_params}}.

\end{itemize}

%% file: 6simulation_results.tex
\section{Simulation Results} \label{simulation_results}

\begin{figure*}[t]
	\setlength{\abovecaptionskip}{3pt plus 3pt minus 2pt}
	\centering
	\includegraphics[width=2\columnwidth]{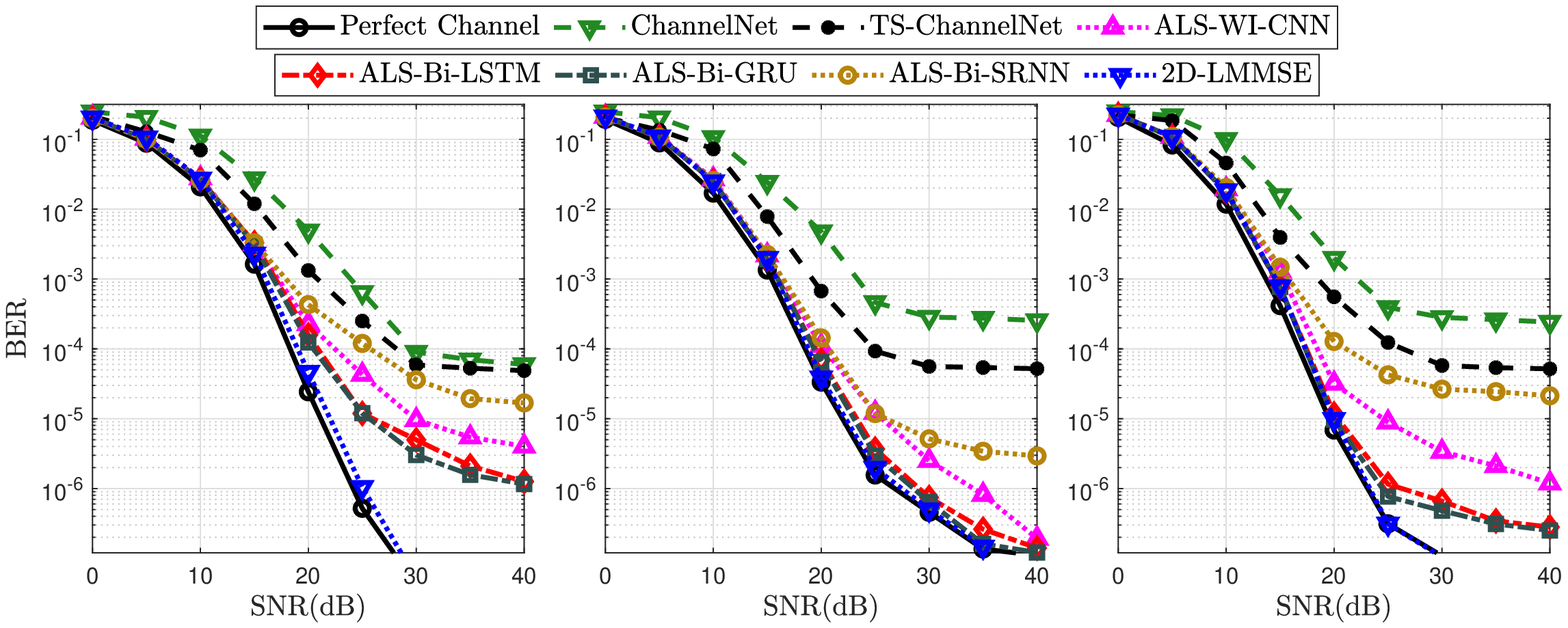}\\[-4ex]
	\subfloat[\label{BER_QPSK_FBF} BER employing QPSK modulation.]{\hspace{.5\linewidth}} \\[-2ex]
	\vspace*{12pt}
	\includegraphics[width=2\columnwidth]{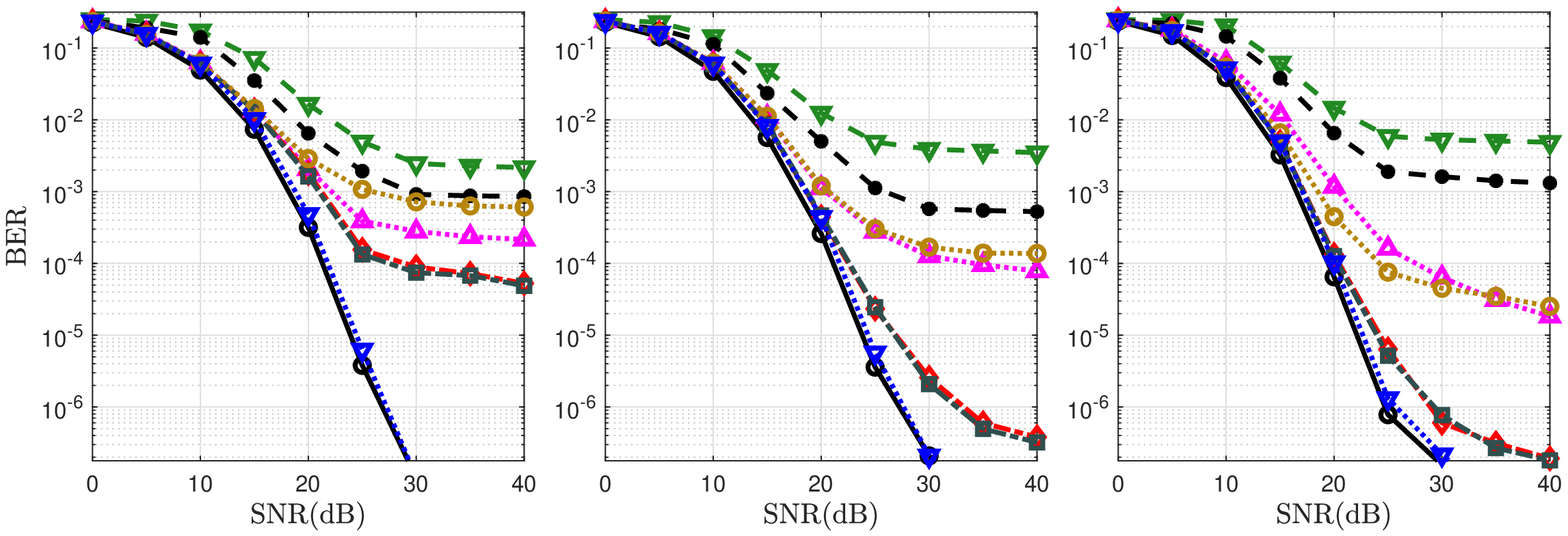} \\[-4ex]
	\subfloat[\label{BER_16QAM_FBF} BER employing 16QAM modulation.]{\hspace{.5\linewidth}} \\
	\caption{BER performance, mobility from left to right: low  ($v = 45~\text{Kmph}, f_{d} = 250$ Hz), high  ($v = 100~\text{Kmph}, f_{d} = 500$ Hz), very high  ($v = 200~\text{Kmph}, f_{d} = 1000$ Hz). The CNN refers to SRCNN and DNCNN in low and high/very high mobility scenarios, respectively.}
\end{figure*}

In this section, the performance evaluations as well as the computational complexity analysis of the CNN-based estimators,  conventional 2D LMMSE estimator, and the proposed Bi-RNN based channel estimator are presented in terms of \ac{BER}. We note that, we only consider the ALS-WI-CNN among the WI-CNN estimators since it has the best performance. It is noted that there exist three main types of Bi-RNNs: (\textit{i}) Bi-Simple RNN (Bi-SRNN), (\textit{ii}) Bi-long short-term memory (Bi-LSTM), and (\textit{iii}) Bi-\ac{GRU}. The Bi-SRNN is used for simple interpolation tasks where the interpolation at each symbol is only affected by the neighboring symbols. However, for longer correlated series, Bi-LSTM and Bi-GRU can be employed, where Bi-GRU provides better performance-complexity trade-off in comparison to other Bi-RNN units. Hence,
the performance of employing Bi-SRNN and Bi-LSTM units instead of the proposed Bi-\ac{GRU} unit are also investigated. 

Vehicular communications are considered as a simulation case study, where three mobility scenarios are defined as: (\textit{i}) low mobility ($v = 45~\text{Kmph}, f_{d} = 250$ Hz) (\textit{ii}) High mobility ($v = 100~\text{Kmph}, f_{d} = 500$ Hz) (\textit{iii}) Very high mobility ($v = 200~\text{Kmph}, f_{d} = 1000$ Hz). The considered OFDM parameters are based on the IEEE 802.11p standard~\cite{ref4}, where $K_{\text{on}} = 52$ active subcarriers. The power-delay profiles of the employed channel models are provided in~\cite{9813719} (Table~$4$). These simulations are implemented using QPSK and 16QAM modulation orders, $I = 100$ as a frame length, $P = \{1, 2, 3\}$ for low, high, and very high mobility scenarios, respectively. The SNR range is $[0 ,5 ,\dots, 40 ]$ dB. In addition, the performance evaluation is made according to the employed modulation order and the mobility scenario.

\subsection{BER Performance}

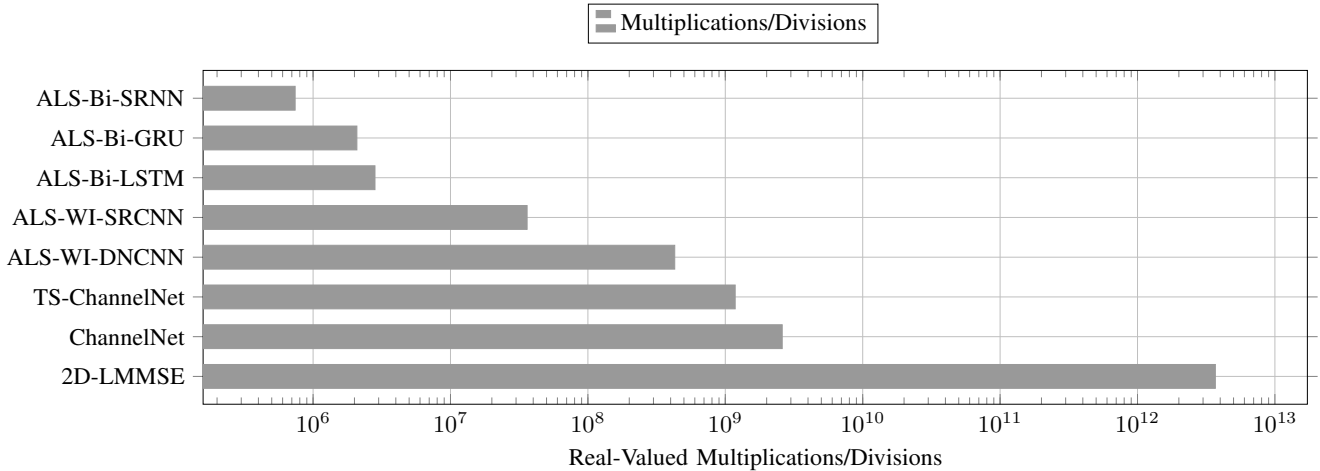
\begin {figure*}[t]
\centering
\scalebox{0.9}{
\begin{tikzpicture}
\begin{axis}[
    xbar,
    xlabel={Real-Valued Multiplications/Divisions},
    symbolic y coords={2D-LMMSE, ChannelNet,TS-ChannelNet,ALS-WI-DNCNN,ALS-WI-SRCNN, ALS-Bi-LSTM, ALS-Bi-GRU, ALS-Bi-SRNN},
    ytick=data,
    xmode=log,
    legend style={at={(0.48,+1.2)},
    anchor=north,legend columns=-1},
    nodes near coords align={horizontal},
    width=2\columnwidth,
    height=6.5cm,
    grid=major,
    cycle list = {gray!80,blue!80,gray!80,blue!80}
    ]
\addplot+[fill] coordinates {  (3686656161000,2D-LMMSE) (2595149600,ChannelNet) (1180150400,TS-ChannelNet) (428595544,ALS-WI-DNCNN) (36108800,ALS-WI-SRCNN) (2821064,ALS-Bi-LSTM) (2083008,ALS-Bi-GRU) (740104,ALS-Bi-SRNN)};
\legend{Multiplications/Divisions} 
\end{axis}
\end{tikzpicture}
}
\caption{Computational complexity comparison of the studied DL-based FBF channel estimators~\cite{9813719}.}
\label{fig:bar_graph_FBF}
\end{figure*}


\subsubsection{Modulation Order}

\figref{BER_QPSK_FBF} and \figref{BER_16QAM_FBF} depict the \ac{BER} performance employing QPSK and 16QAM modulation orders, respectively. The {\ac{ChannelNet}} and {\ac{TS-ChannelNet}} use predefined fixed parameters in the applied interpolation scheme, where the RBF interpolation function and the ADD-TT frequency and time averaging parameters need to be updated in a real-time manner. On the contrary, in the ALS-WI-CNN estimator there are no fixed parameters and the time correlation between the previous and the future pilot symbols is considered in the {\ac{WI}} interpolation operation. These aspects lead to the performance superiority of the ALS-WI-CNN compared to the ChannelNet and TS-ChannelNet estimators.

Although {\ac{CNN}} processing is applied in the {\ac{ChannelNet}}, {\ac{TS-ChannelNet}}, and ALS-WI-CNN estimators, they suffer from a considerable performance degradation that is dominant in very high mobility scenario. This show that the {\ac{CNN}} processing is not able to effectively alleviate the impact of Doppler interference, especially in very high mobility scenarios, where the proposed ALS-Bi-GRU based channel estimation scheme outperforms the WI-ALS-CNN estimator by at least $5$ dB and $12$ dB gain in terms of {\ac{SNR}} for a BER = $10^{-5}$ employing QPSK and 16QAM modulations, respectively. We note that the robustness of the proposed Bi-RNN based channel estimator against high mobility is mainly due to the accuracy of the end-to-end 2D interpolation implemented by the utilized Bi-GRU unit. Moreover, we can see that employing Bi-LSTM performs similar to using Bi-GRU unit in the proposed estimator, this due to the used frame structure, where the variation of the doubly-selective channel within each sub-frame is low. However, it can be noticed that the ALS-WI-CNN performs better than the Bi-SRCNN unit in low and high mobility scenarios, while using Bi-SRCNN unit leads to around $2$ dB gain in terms of {\ac{SNR}} for a BER = $10^{-4}$ in comparison with the ALS-WI-CNN estimator in very high mobility scenario as shown in~\figref{BER_16QAM_FBF}.

As a result, we can conclude that employing Bi-GRU unit instead of {\ac{CNN}} network leads to more accurate channel estimation with lower complexity. Finally, we note that the performance of the 2D-LMMSE estimator is comparable to the performance of ideal channel but it requires huge complexity as discussed in Section~\ref{complexity}, which is impractical in real scenario. Moreover, the proposed estimator records almost close performance as the 2D-LMMSE estimator. Therefore, the proposed ALS-Bi-GRU based channel estimator is an alternative to the 2D-LMMSE estimator where it provides a good performance-complexity trade-off.

\subsubsection{Mobility}

The impact of mobility can be clearly observed in \figref{BER_16QAM_FBF}, where the performance of the the {\ac{ChannelNet}} and {\ac{TS-ChannelNet}} channel estimation schemes degrades as the mobility increases, and the impact of the time diversity gain is not dominant due to the high estimation error of the 2D RBF and ADD-TT interpolation techniques employed in the {\ac{ChannelNet}} and {\ac{TS-ChannelNet}} estimators, respectively. In contrast, the time diversity gain is dominant in the ALS-WI-CNN and the proposed ALS-Bi-GRU channel estimator, since the ALS and WI estimations are accurate, thus,  the SR-CNN and DN-CNN networks are capable of overcoming the Doppler interference. However, using the {\ac{ALS}} estimation at the pilot symbols followed by Bi-GRU unit for 2D interpolation at the data symbols reveal a considerable robustness against mobility. This is due to the ability of the optimized Bi-GRU unit in significantly alleviating the impact of Doppler interference, where it can be noticed that the proposed estimator is able to outperform the ALS-WI-CNN estimators in different mobility scenarios. 


\subsection{Computational Complexity Analysis \label{complexity}}

This section provides a detailed computational complexity analysis of the studied channel estimation schemes. The computational complexity analysis is performed in accordance with the number of real-valued arithmetic multiplications/divisions necessary to estimate the channel for one received {\ac{OFDM}} frame~\cite{9813719}.

The computational complexity of any Bi-RNN unit is twice the required complexity for the regular RNN unit, since both forward and backward data flow are applied. According to~\cite{Li2017ReducingTC}, the overall computational complexity required by Bi-SRNN, Bi-LSTM, and Bi-GRU can be expressed by $2Q K_{in} + 4Q^{2}$,  $8Q K_{in} + 8Q^{2} + 6Q$, and $6Q K_{in} + 6 Q^{2} + 6Q$ multiplications/divisions, respectively, where $Q$ denotes the RNN hidden size. 
The proposed Bi-GRU estimator is optimized where $Q = 32$. Moreover, we use $P = 3$, i.e. assuming very high mobility scenario, in order to have fair comparison with the ALS-WI-CNN estimator, and $K_{in} = 2 K_{\text{on}} I$. The {\ac{ALS}} channel estimation at the inserted pilot symbols requires $4 K^{2}_{\text{on}} P + 2 K_{\text{on}} P + 2 K_{\text{on}}$ multiplications/divisions. Therefore, the overall computational complexity of the proposed Bi-{\ac{GRU}} channel estimation scheme can be expressed by $16 K^{2}_{\text{on}}  + 39946 K_{\text{on}} + 6336$ multiplication/divisions. 

We note that employing Bi-LSTM instead of the GRU unit increases the computational complexity by around $26.29\%$ where  $16 K^{2}_{\text{on}} + 53258 K_{\text{on}} + 8384$
multiplications/divisions are needed without any gain in the BER performance as discussed in~\ref{simulation_results}. Moreover, using Bi-SRNN requires $16 K^{2}_{\text{on}} + 13322 K_{\text{on}} + 4096$ multiplications/divisions. Therefore, the overall computational complexity is decreased by  $73.63\%$ and $64.22\%$ in comparison to the ALS-Bi-LSTM and ALS-Bi-GRU estimators, respectively. However, Bi-SRNN unit suffers from limited performance due to its simple architecture.

\figref{fig:bar_graph_FBF} illustrates the computational complexities of the studied CNN-based FBF channel estimators. We can notice that the conventional 2D LMMSE estimator records the highest computational complexity~\cite{ref_LMMSE_Computational}, making it impractical in real-time scenarios. Moreover, the ChannelNet, {\ac{TS-ChannelNet}}, and the WI-CNN estimators did not provide a good complexity vs. performance trade-off. In contrast, the complexity is significantly decreased by the proposed ALS-Bi-GRU channel estimator where it is $10$x and $115$x times less complex than the ALS-WI-SRCNN and the ALS-WI-DNCNN estimators, respectively. Moreover, the proposed ALS-Bi-GRU channel estimator is $10^{6}$x less complex than the conventional 2D LMMSE channel estimator. Therefore, we can conclude that employing the optimized Bi-RNN networks instead of CNN networks in the channel estimation is more feasible and at the same time it offers better performance due to the good interpolation ability of the Bi-RNN networks. Thus making the proposed estimator a good alternative to the 2D LMMSE as well as the CNN-based channel estimation schemes. 

%% file: 7conclusions.tex
\section{Conclusion} \label{conclusions}
In this paper, {\ac{DL}}-based FBF channel estimation in doubly-selective environments is studied. The recently proposed CNN-based channel estimators have been extensively surveyed, where their limitations besides the drawbacks of the conventional 2D LMMSE estimator are presented. In this context, we have proposed a low-complexity, robust, and adaptive {\ac{Bi-RNN}} based channel estimation scheme, where the great potential of Bi-GRU in performing end-to-end 2D interpolation is exploited. Unlike the recently proposed ChannelNet, TS-ChannelNet, and {\ac{WI}}-{\ac{CNN}} estimators that suffer from high computational complexity and performance degradation in high mobility vehicular scenarios, the proposed {\ac{Bi-RNN}} based estimator have substantially reduced the computational complexity, while recording a significant performance superiority over the {\ac{SoA}} CNN-based channel estimators in all mobility scenarios. Moreover, the proposed estimator is less complex than the conventional 2D LMMSE estimator by at least $10^{6}$ times while recording a convenient BER performance especially in high mobility scenarios, which makes it a good alternative to the conventional 2D LMMSE estimator.